\documentclass[%
reprint,
showpacs,
 amsmath,amssymb,
 aps,prc,
superscriptaddress,
]{revtex4-2}

\usepackage{siunitx}
\usepackage{graphicx}
\usepackage{dcolumn}
\usepackage{bm}
\usepackage{isotope}
\usepackage{hyperref}
\usepackage{booktabs}
\usepackage{multirow}

\newcolumntype{C}[1]{>{\centering\arraybackslash}m{#1}}
\begin{document}


\title{Determination of $^{170,172}$Yb($\alpha,n$)$^{173,175}$Hf reaction cross sections in a stacked-target experiment}

\author{M.~M\"uller}
\email[]{mmueller@ikp.uni-koeln.de}
\affiliation{University of Cologne, Institute for Nuclear Physics,  50937 Cologne, Germany}
\author{F.~Heim}
\affiliation{University of Cologne, Institute for Nuclear Physics,  50937 Cologne, Germany}
\author{Y.~Wang}
\affiliation{University of Cologne, Institute for Nuclear Physics,  50937 Cologne, Germany}
\author{S.~Wilden}
\affiliation{University of Cologne, Institute for Nuclear Physics,  50937 Cologne, Germany}
\author{A.~Zilges}
\affiliation{University of Cologne, Institute for Nuclear Physics,  50937 Cologne, Germany}

\date{24 January 2023}
             
\begin{abstract}
\setlength{\parindent}{0pt}

\textbf{Background:} For understanding the synthesis of elements in the universe, precise knowledge of reaction rates and cross sections is paramount.
This is especially true for the $p$ process as its study requires large network calculations including thousands of nuclei and nuclear reactions with little room for simplification.
Therefore, robust theoretical methods for predicting cross sections are needed which are usually based on Hauser-Feshbach calculations.
These calculations use physics input in the form of $\gamma$-ray strength functions, nuclear level densities and particle+nucleus optical-model potentials.
These have to be constrained using experimental results.

\textbf{Purpose:} To constrain the $\alpha$ optical-model potential $\alpha$-induced reactions are well-suited.
The ytterbium isotopic chain not only offers multiple stable isotopes on which cross sections can be measured and insights into the evolution of the $\alpha$ optical-model potential with the neutron-to-proton ratio can be gained but also includes the $p$ nucleus $^{168}$Yb. 
Its abundance is significantly impacted by the $^{164,166}$Yb($\alpha,\gamma$) reactions and is, therefore, also affected by the constraints resulting from the experiment presented in this paper.

\textbf{Method:} In order to study the $^{170,172}$Yb($\alpha,n$)$^{173,175}$Hf reaction cross sections the activation method was used.
During irradiation the targets were arranged in stacks of four to reduce the required irradiation time.
Aluminum degrader foils served as backings.
The average interaction energy inside each ytterbium layer was determined via \textsc{Geant4} simulations.
A third layer, consisting of manganese, was used to verify the simulations by comparing the measured $^{55}$Mn($\alpha,(2)n$)$^{57,58}$Co reaction cross sections to previous results.
For irradiation the \mbox{10 MV FN tandem} accelerator located at the University of Cologne was used and the activation measurement was performed utilizing the \textsc{Cologne Clover Counting} setup consisting of two clover-type high-purity germanium detectors in a face-to-face geometry.

\textbf{Results:} For the $^{170}$Yb($\alpha,n$) reaction seven cross sections at center-of-mass energies between 12.7 and 16.5 MeV were measured.
For the $^{172}$Yb($\alpha,n$) reaction six cross sections for center-of-mass energies of 13.1 to \mbox{16.5 MeV} could be determined with an additional upper limit at \mbox{E$_{c.m.}$ = 12.3 MeV}.

\textbf{Conclusion:} Comparisons to theoretical models show that state-of-the-art $\alpha$-optical model potentials are able to reproduce the measured cross sections very well.
The ratios of ($\alpha, n$) reaction cross sections in the ytterbium isotopic chain can be accurately reproduced as well.

\end{abstract}

\maketitle

\section{Introduction}
While almost all elements heavier than iron are produced in neutron-capture reaction networks, there are 35 neutron deficient so-called $p$ nuclei, for which this is not the case \citep{B2FH}.
Multiple production mechanisms for $p$ nuclei such as the $\gamma$, $rp$, $\nu p$ and $np$ processes have been suggested, with the $\gamma$ process dominating the production of higher mass $p$ nuclei like $^{168}$Yb \cite{rp-process,nup-process,np-process,pprocess_review}.
The $\gamma$ process is based on photodisintegration reactions on heavy seed nuclei produced in neutron-capture processes.
It is most likely to take place in scenarios, such as type II supernovae, at temperatures around 3 GK \cite{ptemp,Pignatari_2016}.
In environments, in which the $\gamma$ process is possible, the ($\gamma,n$), ($\gamma,p$) and ($\gamma, \alpha$) photodisintegration reactions not only compete with each other, but also with their inverse reactions and $\beta$ decay \cite{Woosley_1978}.
This results in a process path which is very spread out and therefore necessitates detailed theoretical modeling capable of calculating all of the reaction cross sections involved within the astrophysically relevant Gamow window. \\

The framework in which such reaction cross sections are calculated is usually the Hauser-Feshbach theory, which requires nuclear physics input in the form of transmission coefficients and nuclear level information \cite{HauserFeshbach}.
This nuclear input can in turn be calculated using particle-nucleus optical-model potentials (OMP), $\gamma$-ray strength functions ($\gamma$-SF), and nuclear level densities (NLD).
The Hauser-Feshbach theory can then be used to provide reaction rates for network calculations \cite{HauserFeshbach}.
Many different models are available to calculate these quantities, but all of them have to be adjusted to and verified by experimental data.
What kind of experiment is best suited for this kind of validation strongly depends on which nuclear input parameter is under investigation and a wealth of methods is available to deduce the needed information \cite{Rauscher_2013}. \\

The methods used at the University of Cologne mostly depend on the determination of total and partial reaction cross sections \cite{Heim_2020_technique}.
In recent years, investigations of the $\gamma$-SF and the NLD have been performed using radiative proton-capture experiments at the \textsc{Horus} $\gamma$-ray spectrometer for in-beam as well as the \textsc{Cologne Clover Counting} setup for activation measurements \cite{Horus,Clover,Heim_2021_96Mo,Heim_2021_109Ag,Heim_2021_94Mo,Heim_2020_94Mo,Scholz_2020_SF}.
The same experimental setups have also been used in the determination of cross sections of $\alpha$-induced reactions for investigations of the $\alpha$-OMP \cite{Scholz_2016_108Cd,Netterdon_2015_112Sn}. \\

\begin{figure}[tb]
\centering
\includegraphics[width=\columnwidth, keepaspectratio]{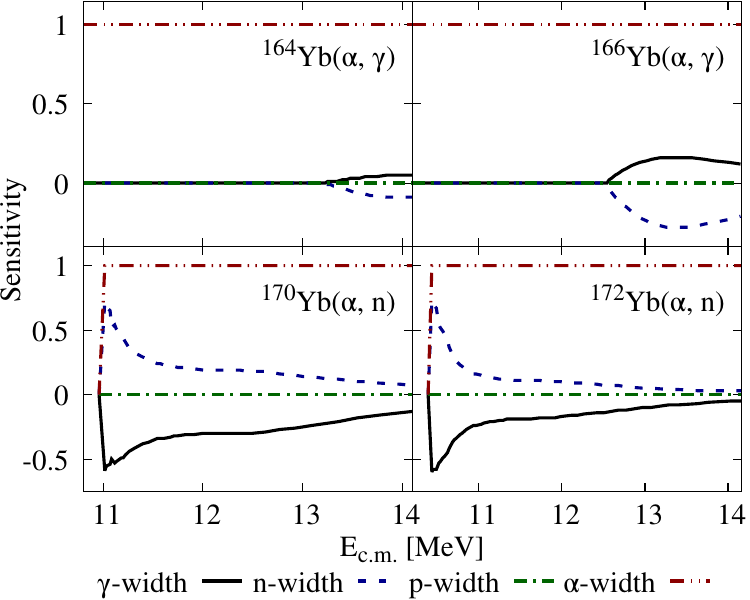}
\caption{Sensitivity of theoretical cross sections for the $^{164,166,170,172}$Yb($\alpha,n$)$^{167,169,173,175}$Hf reactions to the $\alpha$-, \mbox{$\gamma$-,} proton- and neutron-widths. The sensitivities were taken from Ref. \cite{Rauscher}.}
\label{fig:sens}
\end{figure}

By systematically varying the reaction rates that go into network calculations, the reactions that contribute the most to the production uncertainties of $p$ nuclei can be identified.
Varying the transmission coefficients themselves can narrow down which quantity needs to be investigated to reduce uncertainties in theoretical calculations even further.
Such studies by Rauscher and Nishimura indicate, that for $^{168}$Yb the $^{164,166}$Yb($\alpha, \gamma$)$^{168,170}$Hf reactions are of key importance and that both reactions are mainly sensitive to the $\alpha$-width and therefore to the $\alpha$-OMP \cite{Rau-key,Nis-key}.
These reactions, however, cannot be measured directly as $^{164}$Yb and $^{166}$Yb are unstable.
Instead, the $^{168}$Yb($\alpha, \gamma$)$^{172}$Hf and $^{168}$Yb($\alpha, n$)$^{171}$Hf reactions were measured by Netterdon \textit{et al.} because $^{168}$Yb is the lightest stable Yb isotope \cite{Netterdon168Yb}.
Constraints on these reactions can be used to lower the uncertainties in the key reaction rates via extrapolation.
This extrapolation can be improved by further investigating the evolution of the $\alpha$-OMP with the proton-to-neutron ratio in the Yb isotopic chain.
For this purpose the $^{170,172}$Yb($\alpha,n$)$^{173,175}$Hf reactions were investigated which, as Fig. \ref{fig:sens} shows, are also mostly sensitive to the $\alpha$-OMP \cite{Rauscher}.

\section{Experimental Setup and method}
\label{sec:Experimental Setup}

\begin{figure}[tb]
\centering
\includegraphics[width=\columnwidth, keepaspectratio]{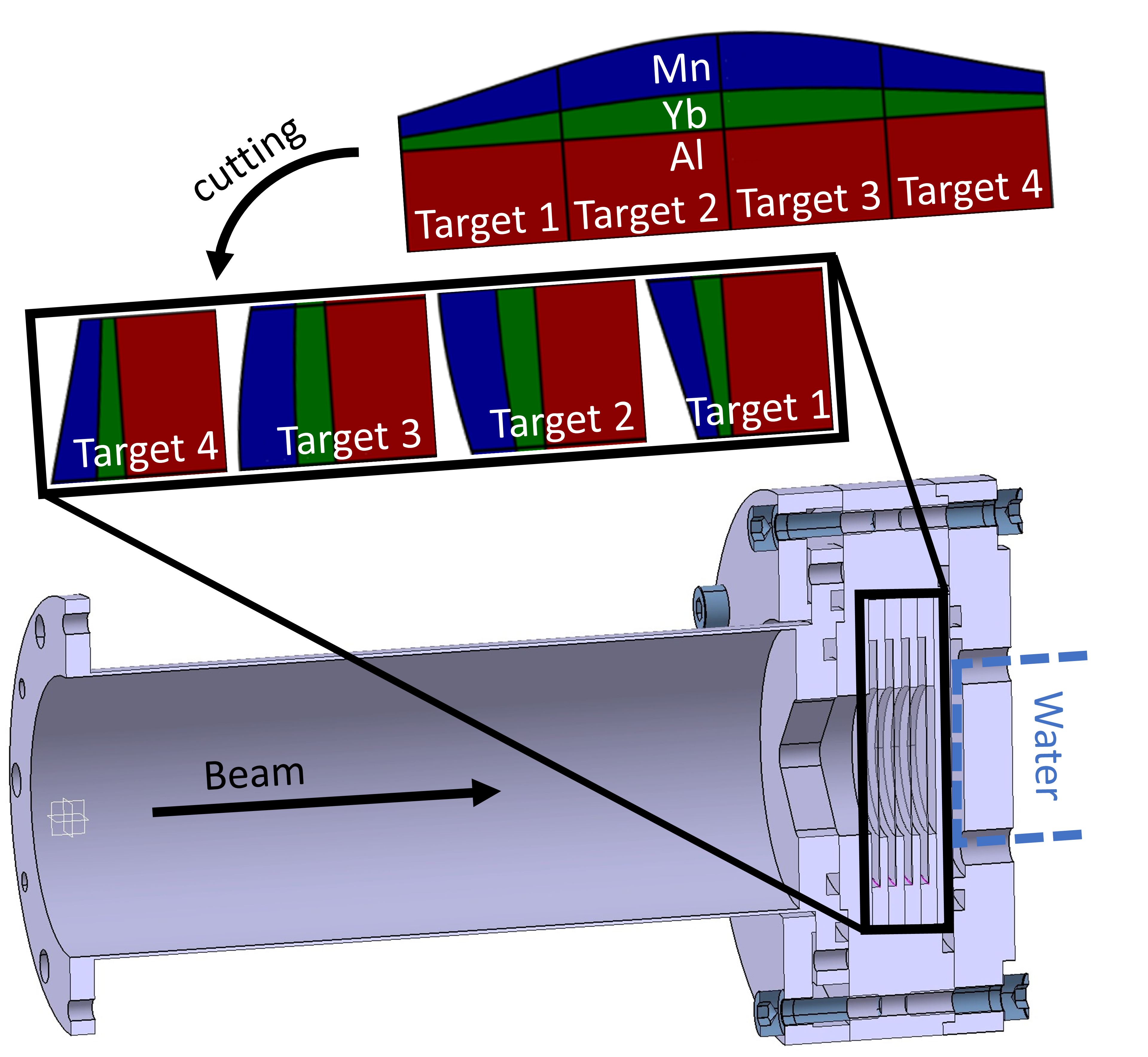}
\caption{Water cooled target chamber designed to accommodate a stack of four targets.
Above the chamber the target composition and the arrangement of targets in a stack are illustrated.}
\label{fig:chamber}
\end{figure}

The experiment was performed utilizing the activation technique \cite{Gyuerky2019}.
To facilitate the application of the stacked-target method a new water-cooled chamber, capable of holding up to four targets, was constructed (see Fig. \ref{fig:chamber}).
After irradiation, these targets were transferred to the \textsc{Cologne Clover Counting} setup consisting of two clover-type high-purity germanium detectors arranged in a face-to-face geometry for $\gamma$-ray detection.
The distance between the targets and the detectors was 13 mm which resulted in a total detection efficiency of about 5 \% at \mbox{E$_{\gamma}$ = 1.3 MeV}.
Depending on the yield, targets were measured for up to 10 days.
Counting times as well as the waiting times that past between the end of irradiation and the start of a measurement are shown in Tab. \ref{tab:targets}.
The following sections will provide detailed information about the target composition and the stopping process of the beam inside of each target layer.

\subsection{Target composition}

\begin{table*}
\begin{center}
\caption[Targets]{Interaction energies (E$_{c.m.}$) and areal densities (N$_T$) as well as activation (t$_A$), waiting (t$_W$), and counting (t$_C$) times for all 16 targets used in the experiment.}
\label{tab:targets}
\begin{tabular}{cccccccc|cccccccc}
\\
\multicolumn{2}{c}{$^{55}$Mn} & \multicolumn{2}{c}{$^{170}$Yb} & $^{27}$Al & \multicolumn{3}{c}{} & \multicolumn{2}{c}{$^{55}$Mn} & \multicolumn{2}{c}{$^{172}$Yb} & $^{27}$Al & \multicolumn{3}{c}{} \\
E$_{c.m.}$ & N$_T$ & E$_{c.m.}$ & N$_T$ & E$_{c.m.}$ & t$_A$ & t$_W$ & t$_C$ & E$_{c.m.}$ & N$_T$ & E$_{c.m.}$ & N$_T$ & E$_{c.m.}$ & t$_A$ & t$_W$ & t$_C$ \\
{[}MeV] & [$\frac{mg}{cm^2}$] & [MeV] & [$\frac{mg}{cm^2}$] & [MeV] & [h] & [h] & [h] & [MeV] & [$\frac{mg}{cm^2}$] & [MeV] & [$\frac{mg}{cm^2}$] & [MeV] & [h] & [h] & [h] \\ [0.5 mm] \hline \multicolumn{7}{c}{} & & \multicolumn{8}{c}{} \\ [-2ex]
11.5(1) & 0.41(4) & 12.0(1) & 0.18(2) & 10.0(3) & 45.6 & 748.3 & 56.0 & 10.9(2) & 0.76(8) & 11.5(1) & 0.46(5) & 9.3(3) & 70.9 & 781.9 & 44.8 \\
12.1(1) & 0.37(4) & 12.7(1) & 0.20(2) & 10.5(3) & 45.6 & 5.4 & 70.3 & 11.7(1) & 0.72(7) & 12.3(1) & 0.40(4) & 10.1(4) & 70.9 & 844.7 & 198.0 \\
12.78(7) & 0.36(4) & 13.41(6) & 0.20(2) & 11.2(2) & 45.6 & 2.1 & 3.2 & 12.5(1) & 0.79(8) & 13.13(8) & 0.35(4) & 11.0(2) & 70.9 & 104.8 & 245.2 \\
13.40(4) & 0.37(4) & 14.06(2) & 0.18(2) & 11.8(3) & 45.6 & 0.4 & 1.7 & 13.24(9) & 0.85(8) & 13.92(3) & 0.41(4) & 11.6(3) & 70.9 & 53.3 & 46.6 \\
13.9(1) & 0.45(4) & 14.6(1) & 0.31(3) & 12.3(3) & 20.8 & 6.5 & 16.3 & 13.8(1) & 0.65(6) & 14.5(1) & 0.33(3) & 12.3(3) & 21.5 & 44.8 & 24.3 \\
14.54(9) & 0.45(4) & 15.25(9) & 0.34(3) & 12.9(2) & 20.8 & 2.7 & 3.7 & 14.5(1) & 0.60(6) & 15.19(9) & 0.32(3) & 12.9(3) & 21.5 & 6.8 & 15.4 \\
15.12(8) & 0.46(5) & 15.88(6) & 0.29(3) & 13.5(2) & 20.8 & 1.6 & 1.0 & 15.08(8) & 0.55(6) & 15.83(6) & 0.29(3) & 13.5(2) & 21.5 & 4.0 & 2.2 \\
15.71(5) & 0.50(5) & 16.48(2) & 0.29(3) & 14.1(2) & 20.8 & 0.5 & 1.0 & 15.69(5) & 0.57(6) & 16.46(2) & 0.26(3) & 14.1(2) & 21.5 & 3.0 & 1.0 \\
\end{tabular}
\end{center}
\end{table*}

For the experiment two types of targets were produced in the institute's target laboratory.
For the first type a layer of ytterbium enriched to a $^{170}$Yb content of \mbox{83.2(3) \%} was deposited onto a large piece of pure aluminum foil by evaporation.
On top of that layer a second layer of $^{55}$Mn was deposited, also via evaporation.
The targets used in the experiment were then cut from this larger piece of foil (see upper part of Fig. \ref{fig:chamber}).
The second type of targets was produced in the same way but this time using ytterbium material with a $^{172}$Yb enrichment of 97.1(1) \%.
Evaporation deposition produces a Gaussian profile.
Covering a larger area increases the width of the Gaussian distribution and therefore produces targets with a more uniform thickness.
It also reduces the material loss compared to producing each target individually.
A sketch of the targets is shown in Fig. \ref{fig:chamber}.
In order to determine the precise areal densities of each target layer a Rutherford-Backscattering (RBS) experiment was performed at the RUBION facility of the Ruhr University Bochum.
The results are displayed in Tab. \ref{tab:targets}. \\

Note, that the manganese layer is used for validation of the experimental results through the well known $^{55}$Mn($\alpha, (2)n$)$^{57,58}$Co reaction cross sections and was deposited on top of the ytterbium layer instead of producing separate manganese targets in order to prevent oxidation of the ytterbium.

\subsection{Beam properties}
During the experiment four stacks of four targets each, two consisting of $^{170}$Yb targets and two of $^{172}$Yb targets, were irradiated with $\alpha$-particles.
The two stacks per Yb isotope were irradiated at beam energies of \mbox{14.5 MeV} and 17 MeV, respectively, with a beam current of roughly 200 nA.
Currents were recorded using a current integrator attached to the chamber at 0.2 s intervals.
As the chamber conducts electricity throughout, this measurement automatically takes care of $\delta$ electrons and no suppression voltage had to be applied.
However, this also means, that the entire beam does not necessarily have to hit the targets, but might also partially hit the walls of the chamber without changing the measured current.
This is one of the reasons, why validation of the results via the manganese reactions is imperative. \\

The most important reason for using a benchmark reaction, however, is that the interaction energy inside each layer of each target in the stack needs to be determined.
This was accomplished using a \textsc{Geant4} simulation of the energy loss \cite{Geant4}.
The uncertainty of the resulting interaction energy within each layer stems from three factors: the width of the energy distribution obtained in the simulation, the energy loss in the layer itself, and the uncertainty of the areal target densities.
The influence of the latter was determined by repeating the \textsc{Geant4} simulation 10$^4$ times while randomly varying the areal densities within their uncertainties.
The energies and their uncertainties resulting from this simulation as well as the irradiation times for each target can be found in Tab. \ref{tab:targets}.
As this was the first time this specific simulation was used, its results were verified not only by comparison of the reference reactions to previous results but also by comparison to \textsc{Srim} simulations \cite{SRIM}.
The \textsc{Srim} simulations corroborated the \textsc{Geant4} simulations. \\

The simulation shows, that with a beam energy of \mbox{14.5 MeV} energies very close to the astophysically relevant Gamow window located between 7.9 and 11.3 MeV for a temperature of T = 3 GK were reached \cite{Rauscher_gamow}.
Note, that independent of the experimental setup's sensitivity, activation measurements deeper into the Gamow window are impossible due to the energy thresholds for the $^{170,172}$Yb($\alpha, n$) reactions at about \mbox{11.2 MeV} and \mbox{10.5 MeV} and the fact, that the $^{170,172}$Yb($\alpha, \gamma$) reactions produce stable nuclei.

\section{Data Analysis}
\begin{figure}[tb]
\centering
\includegraphics[width=\columnwidth, keepaspectratio]{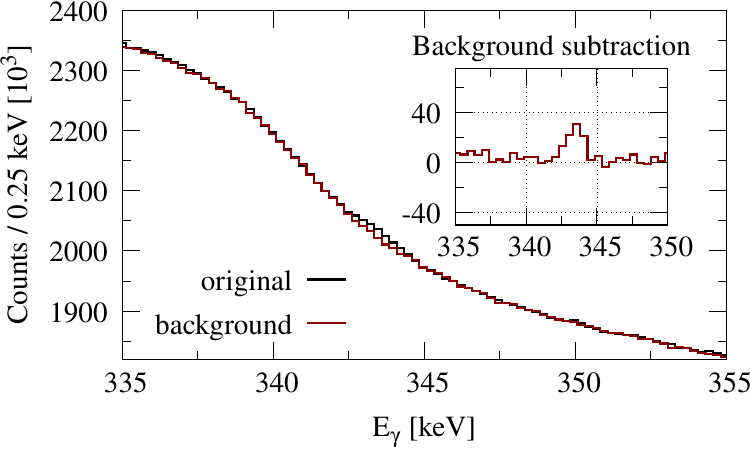}
\caption{Normalization of a background spectrum to a spectrum obtained from a $^{172}$Yb target.
The original spectrum was obtained at an energy of E$_{c.m.}$ = 13.9 MeV and the background spectrum at an energy of E$_{c.m.}$ = 12.0 MeV.
The inset shows the background subtraction resulting from it.}
\label{fig:bgs}
\end{figure}

\begin{figure*}[tb]
\centering
\begin{minipage}{\columnwidth}
\includegraphics[width=\columnwidth, keepaspectratio]{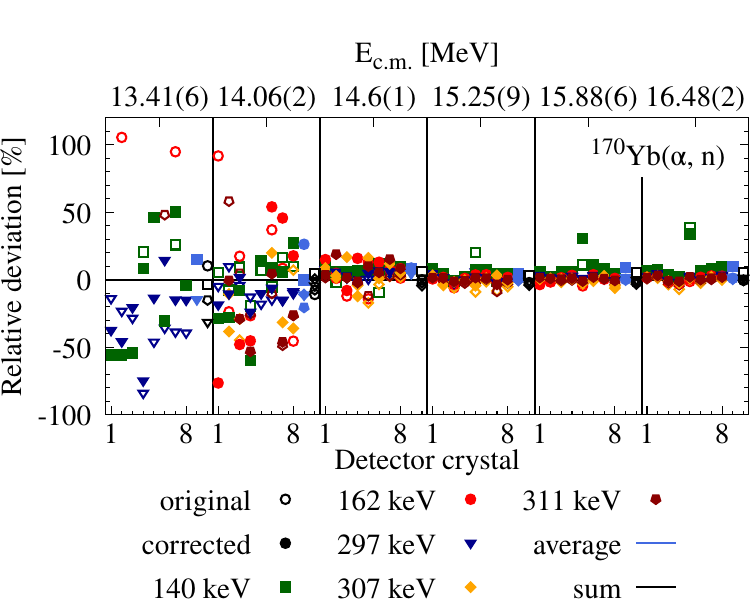}
\end{minipage}
\hfill
\begin{minipage}{\columnwidth}
\includegraphics[width=\columnwidth, keepaspectratio]{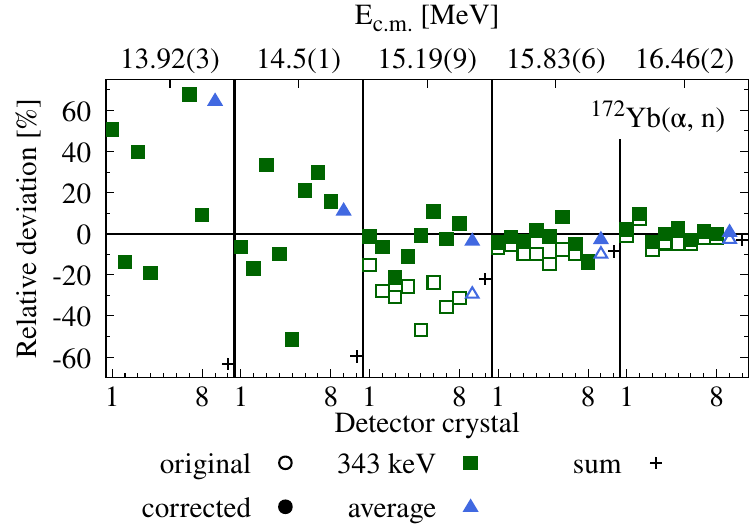}
\end{minipage}
\caption{Relative deviations between reaction yields determined from the spectra of all eight detector crystals and all observed $\gamma$-ray lines.
Which $\gamma$-ray line was used is coded in colors and symbol shapes.
The lower x-axis shows the number of the detector crystal used, where number 9 corresponds to the average of all individual detector crystals (light blue) and number 10 to the yield determined from the sum spectra (black).
Whether the reaction yield was determined from original or background corrected spectra is marked by open or closed symbols.
Yields determined from background corrected sum spectra were used as reference values.
The upper x-axis displays the interaction energy at which the reaction yields were measured with vertical lines separating individual targets.}
\label{fig:Yb}
\end{figure*}

\begin{table}
\begin{center}
\caption{Half-lives and decay radiation information for the nuclei produced in the experiment.
Values were taken from Refs. \cite{nds57,nds58,nds173,nds175}.}
\begin{tabular}{cccc}
\\
Nucleus & T$_{1/2}$ [h] & E$_{\gamma}$ [keV] & rel. $\gamma$-ray intensity [\%] \\ \hline \\ [-2ex]
$^{57}$Co & 6521.8(14) &	122.0607(1) & 85.6(2) \\
 &  &	136.4736(3) & 10.68(8) \\
$^{58}$Co & 1700.6(14) &	810.759(2) & 99.5(2) \\
 &  &	863.95(6) & 0.686(2) \\
 &  &	1674.725(7) & 0.516(1) \\
$^{173}$Hf & 23.6(1) &	123.68(2) & 83(4) \\
 &  &	134.96(1) & 4.7(2) \\
 &  &	139.63(2) & 12.7(6) \\
 &  &	162.01(2) & 6.5(3) \\
 &  &	296.97(2) & 33.9(14) \\
 &  &	306.57(2) & 6.4(3) \\
 &  &	311.24(2) & 10.7(4) \\
$^{175}$Hf & 1680(48) &	343.40(8) & 84.0(3) \\
\label{tab:lines}
\end{tabular}
\end{center}
\end{table}

Despite the importance of the $^{55}$Mn reactions measured in parallel as a benchmark, they also introduced some problematic background.
Most importantly there is only one observable $\gamma$-ray line stemming from the $^{175}$Hf decay and it is located at the Compton edge stemming from the 511 keV line, which, while unavoidable, is also heavily produced in the decay of $^{58}$Co.
For low reaction energies this is a problem, as it becomes almost impossible to correctly determine the number of 343 keV events stemming from the $^{175}$Hf decay.
Information on the half-lives and the decay radiation of reaction products involved can be found in Tab. \ref{tab:lines}.
The problem was solved by performing a background subtraction, that will be discussed in the following subsections.

\subsection{Background subtraction}

For the background subtraction a $^{170}$Yb target irradiated at an energy low enough to not exhibit any $^{173}$Hf decay lines was used.
This precludes the existence of $^{175}$Hf events in the background spectrum stemming from reactions on $^{172}$Yb contaminations in the $^{170}$Yb enriched material.
Figure \ref{fig:bgs} shows an example of such a background subtraction. \\

The background subtraction was performed bin-wise and therefore the same was done for the uncertainty propagation:

\begin{equation}
\begin{split}
N_i &= N_{orig,i} - f \cdot N_{back,i} \\
\Delta N_i &= \sqrt{N_{orig,i} + f^2 \cdot N_{back,i}}
\end{split}
\end{equation}

Here N$_i$, N$_{orig,i}$, and N$_{back,i}$ are the number of counts in bin i of the background corrected, the original, and the background spectrum.
The factor f was determined by normalizing the background spectrum to the two strongest lines in the original spectra, which are the \mbox{511 keV} and the \mbox{811 keV} lines.
As the peak volume corresponds to a sum of bins the uncertainty was calculated as:

\begin{equation}
\Delta N_{\gamma} = \sqrt{\sum_{-3 \sigma \leq i \leq 3 \sigma} \Delta N_{orig,i} + f^2 \cdot N_{back,i}}
\end{equation}

Here N$_{\gamma}$ is the total amount of counts in a peak, and $\sigma$ is the width of the Gaussian distribution fit to the peak in units of bins.
Using the background subtraction peaks previously indistinguishable from the background could be fit, leading to an increase in sensitivity by about a factor of four. \\

Due to the power of the above procedure in reducing the background, it was also applied to spectra taken with $^{170}$Yb targets using a background spectrum obtained from a $^{172}$Yb target.

\subsection{Validation}
The background subtraction was validated by comparing the yields determined from background corrected and original spectra at energies for which the statistics are good enough to distinguish peaks from the background.
Multiple ways of determining the reaction yield are available:
\begin{itemize}
\item Determining the yield from each individual detector crystal's spectrum
\item Averaging over the yields determined from each individual detector crystal's spectrum
\item Adding up the spectra of the individual detector crystals and determining the yield from it
\item Using the original spectra
\item Using background corrected spectra
\end{itemize}
A comparison of the relative deviations between the reaction yields obtained in all of these ways can be found in Fig. \ref{fig:Yb}.
Here, yields determined from the background corrected sum spectra were used as reference values, as this is the only method applicable to the lowest energies.
Note, that energies at which this is the only applicable method are not shown.
One can clearly see that as long as statistics are sufficient, all four methods of determining yields are in very good agreement.
As energies and statistics get lower, the spread of reaction yields gets larger and yields determined from original spectra tend to get smaller than the reference cross sections (negative deviation).
This is to be expected, as for lower statistics, a higher relative amount of actual events is masked by the background.

\section{Results}
\label{sec:results}

\begin{figure*}[tb]
\centering
\includegraphics[width=\linewidth, keepaspectratio]{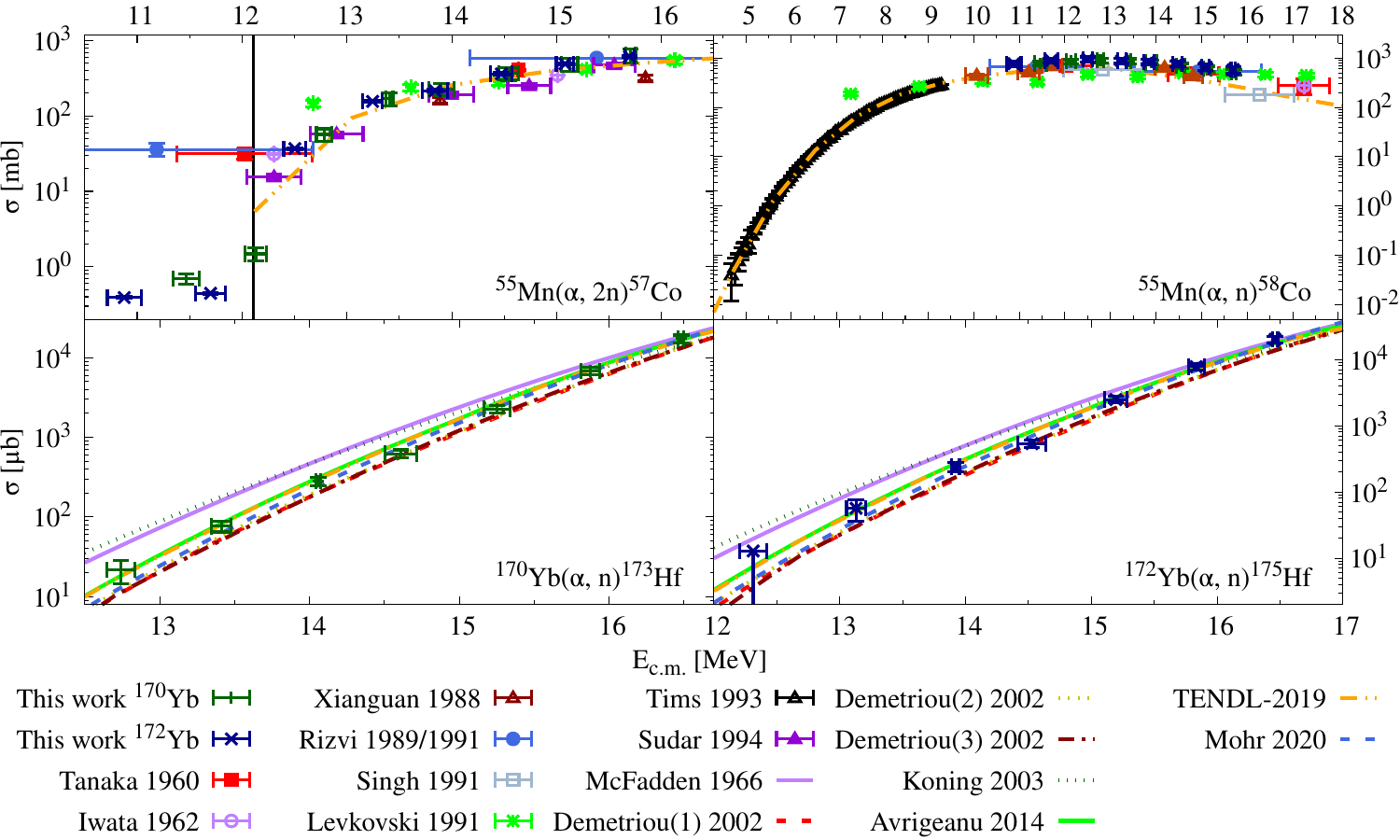}
\caption{Reaction cross sections determined for the $^{55}$Mn($\alpha,(2)n$)$^{57,58}$Co reference reactions (top panels) and the $^{170,172}$Yb($\alpha,n$)$^{173,175}$Hf reactions of interest (bottom panels).
For the reference reactions previous results as well as theoretical values taken from the \textsc{Tendl-2019} database are shown \cite{tanaka,Iwata,Xianguan,Rizvi,Singh,Levkovski,Tims,Sudar, TENDL-2019}.
Which kind of target was used to obtain the reaction cross sections is indicated by the labels $^{170}$Yb and $^{172}$Yb.
More information on the analysis of the reference reactions will be published in a forthcoming paper.
For the Yb reactions theoretical calculations performed using the \textsc{Talys1.95} code are shown \cite{talys1.9}.
These calculations utilized various $\alpha$-OMPs introduced in Refs. \cite{aomp-mcfadden, aomp-demetriou, aomp-koning, BTM}.
Three $\alpha$-OMPs by Demetriou $\textit{et al.}$ are shown.
The first of these uses an imaginary part consisting of a volume term only.
The second adds a surface term to the imaginary part and the third one relates the imaginary part consisting of a volume and a surface term to the real part via a disperion relation.
Theoretical values taken from the \textsc{Tendl-2019} database are shown as well.}
\label{fig:all_cross}
\end{figure*}

\begin{figure}[tb]
\centering
\includegraphics[width=\columnwidth, keepaspectratio]{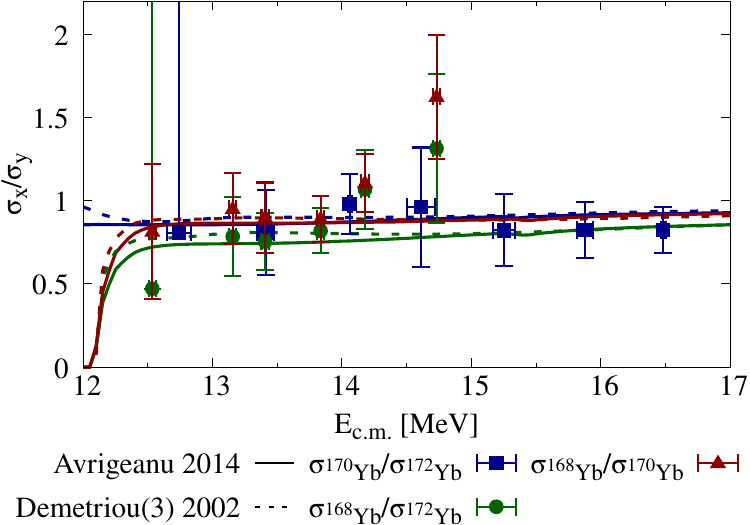}
\caption{Experimental and theoretical ratios of ($\alpha,n$) reaction cross sections in the ytterbium chain.
The ratio of $^{170}$Yb to $^{172}$Yb is shown in blue, $^{168}$Yb to $^{170}$Yb in red, and $^{168}$Yb to $^{172}$Yb in green.
This color code applies to experimental as well as theoretical values.
Theoretical values were calculated with the \textsc{TALYS-1.95} code using the $\alpha$-OMP by Avrigeanu \textit{et al.} (solid) and the dispersive model by Demetriou \textit{et al.} (dashed) \cite{aomp-avrigeanu,aomp-demetriou}.}
\label{fig:ratio}
\end{figure}

\begin{table}
\begin{center}
\caption[Cross sections]{List of absolute $^{170,172}$Yb($\alpha, n$)$^{173,175}$Hf reaction cross sections determined in this work and $^{168}$Yb($\alpha, n$)$^{171}$Hf reaction cross sections taken from Ref. \cite{Netterdon168Yb}.
The value marked with an asterisk is an upper limit.}
\label{Yb_tab}
\begin{tabular}{cccccc}
\\
\multicolumn{2}{c}{$^{168}$Yb($\alpha, n$)} & \multicolumn{2}{c}{$^{170}$Yb($\alpha, n$)}& \multicolumn{2}{c}{$^{172}$Yb($\alpha, n$)} \\
E$_{c.m.}$ & $\sigma$ & E$_{c.m.}$ & $\sigma$ & E$_{c.m.}$ & $\sigma$ \\
{[}MeV] & [$\mu$b] & [MeV] & [$\mu$b] & [MeV] & [$\mu$b] \\ \hline \\ [-2ex]
12.53(3) & 0.012(1) & 12.7(1) & 0.022(7) & 12.3(1) & 0.012(-)* \\
13.15(3) & 0.045(5) & 13.41(6) & 0.08(1) & 13.13(8) & 0.06(2) \\
13.41(3) & 0.068(7) & 14.06(2) & 0.28(4) & 13.92(3) & 0.24(4) \\
13.84(3) & 0.16(2) & 14.6(1) & 0.62(7) & 14.5(1) & 0.54(7) \\
14.18(3) & 0.37(4) & 15.25(9) & 2.3(3) & 15.19(9) & 2.5(3) \\
14.73(3) & 1.3(1) & 15.88(6) & 6.8(8) & 15.83(6) & 7.8(9) \\
- & - & 16.48(2) & 17(2) & 16.46(2) & 20(2) \\
\end{tabular}
\end{center}
\end{table}

For each of the two reference reactions $^{55}$Mn($\alpha,(2)n$)$^{57,58}$Co cross sections were determined for 16 center-of-mass energies between 11.1 and 15.7 MeV.
Good agreement between this and previous experiments can be observed, corroberating the results obtained from the current measurement and the energy loss simulation. \\ 

The lower panels of Fig. \ref{fig:all_cross} display the cross sections measured for the $^{170,172}$Yb($\alpha,n$)$^{173,175}$Hf reactions.
For the $^{170}$Yb($\alpha,n$)$^{173}$Hf reaction seven cross sections could be obtained between 12.7 and 16.5 MeV all of which can be reproduced very well by theory.
The same is true for the six cross sections measured for the $^{172}$Yb($\alpha,n$)$^{175}$Hf reaction as well as the upper limit obtained at a center-of-mass energy of 12.3 MeV.
The upper limit was calculated based on the assumption, that the largest amount of actual events, that could be masked by background is determined by the average background fluctuation.
Using the full-width-at-half-maximum (FWHM) determined at higher statistics, the maximum number of $\gamma$-ray events, that could be hidden by background was determined as the average background fluctuation in a region around the $\gamma$-ray line times the number of bins within the FWHM.
This number of events was used to calculate a cross section, and by adding the uncertainties stemming from the other parameters, that go into the calculation, to it, the upper limit was obtained. \\

In addition to the two ytterbium reactions measured here, cross sections are available for the $^{168}$Yb($\alpha,n$)$^{171}$Hf reaction \cite{Netterdon168Yb}.
This helps with the investigation of the evolution of the $\alpha$-OMP with the neutron-to-proton ratio.
If no experimental data is available to adjust the parameters of a model they have to be chosen based on the assumption, that similar nuclei can be described by similar parameters.
This implies, that the effect of changing a parameter should also be similar for similar nuclei.
However, due to changes in the nuclear structure, adding or taking away even one nucleon can have unexpected effects.
To reproduce these effects, theoretical modeling might require input parameters, that are very different from neighboring nuclei.
To identify which parameters should be changed, the sensitivity S$_{\sigma}$ of a quantity $\sigma$ to a change of a parameter by a factor of f defined by Rauscher as  
\begin{equation}
S_{\sigma} = \frac{\sigma_{new}/\sigma_{old}-1}{f-1}
\end{equation}
should be considered \cite{Rauscher}.
Using this equation the sensitivity of cross section ratios can be written as:
\begin{equation}
S_{\sigma_1/\sigma_2} = \frac{\frac{(\sigma_1/\sigma_2)_{new}}{(\sigma_1/\sigma_2)_{old}}-1}{f-1} = \frac{S_{\sigma_1} - {S_{\sigma_2}}}{S_{\sigma_1}(f-1)+1}
\end{equation}
This shows, that cross section ratios are espacially sensitive to a parameter, if the difference between the two cross sections' sensitivities to that parameter is large.
As sudden changes in the sensitivity point to a change in nuclear structure considering cross section ratios alongside the cross sections themselves can help in identifing key parameters.
Ambiguities in the choice of a parameter set can thus be mitigated. 
Since the experimental cross sections for the three Yb($\alpha, n$) reactions were determined at different energies an exponential interpolation was performed to calculate the cross section ratios.
The procedure applied mixes energy and cross section uncertainties and results in the ratios shown in Fig. \ref{fig:ratio}.

\section{Comparison to theoretical predictions}

The theoretical cross sections shown in Fig. \ref{fig:all_cross} were calculated using the \textsc{Talys1.95} code utilizing various $\alpha$-OMPs \cite{aomp-mcfadden,aomp-koning,aomp-demetriou,aomp-avrigeanu,BTM}.
Aside from choosing different $\alpha$-OMPs, default values were used for all input parameters.
All available NLD and $\gamma$-SF models as well as neutron and proton OMPs were tested as well, but their results do not differ significantly within the considered energy range. \\

The purely phenomenological model by McFadden and Satchler uses a Woods-Saxon shape for both the real and imaginary part of the $\alpha$-OMP with four parameters that can be used to adjust the model to experimental values \cite{aomp-mcfadden, Woods-Saxon}.
These parameters have mostly been constrained through elastic $\alpha$-scattering experiments and the model has been widely used for decades.
However, numerous experiments have since shown, that while the model performs well at high energies, cross sections at sub-Coulomb energies are often overestimated (see e.g. Refs. \cite{Szegedi_2021_100Mo, Kiss_2022_144Sm}).
The same behaviour can be observed here (purple solid). \\

While McFadden and Satchler use volume terms only, more recent phenomenological models like that by Koning \textit{et al.} added surface and spin-orbit contributions as well \cite{aomp-koning}.
This makes the model more flexible, but also increases the complexity and number of parameters that can be adjusted.
Therefore, experimental cross sections can be reproduced more accurately, however, the predictive power remains limited.
Figure \ref{fig:all_cross} shows, that without tuning the input parameters the model by Koning \textit{et al.} also overestimates the cross sections (green dotted). \\

Six theoretical predictions were found to agree very well with the experimental results.
These calculations used five different $\alpha$-OMPs.
Three by Demetriou \textit{et al.}, all of which yield almost the same results within the considered energy range, and the $\alpha$-OMPs by Mohr \textit{et al.} and Avrigeanu \textit{et al.} \cite{aomp-demetriou,aomp-avrigeanu,BTM}.
All five OMPs are global models applicable to a wide range of energies and masses.
The models by Demetriou \textit{et al.} and Mohr \textit{et al.} are semi-microscopic.
They use a double-folding procedure to derive the real part of the OMP from nucleon-nucleon interactions.
Both of these characteristics are paramount to provide the predictive power needed for astrophysical applications in which no experimental cross section data are available.
However, no microscopic derivation of the imaginary part is available.
Instead a phenomenological model usually based on a Woods-Saxon shape is used.
This can be related to the real part obtained in a double folding procedure via the dispersion relation. \\

Demetriou \textit{et al.} offer three parametrizations of the imaginary part all based on the Woods-Saxon shape.
The first one contains a volume term only (red dashed), the second one a volume and a surface term (yellow dotted), and the third one uses the dispersion relation to relate the imaginary part consisting of a volume and a surface term to the real part (dark-red dash-dotted) \cite{aomp-demetriou}. \\

Mohr \textit{et al.} have recently shown, that at sub-Coulomb energies the tail of the imaginary part of the OMP has a very strong influence on cross sections and have proposed an alternative approach to calculating transmission coefficients \cite{BTM}.
The pure barrier transmission model (PBTM) uses the probability to tunnel through the Coulomb barrier to determine transmission coefficients based on the  assumption that at low energies the probability to tunnel through the Coulomb barrier and then back again is negligible.
This potentially leads to a very high predictive power as no free parameters go into this calculation of transmission coefficients.
An accuracy to within a factor of two can be observed here (blue dashed) as has been claimed by Mohr \textit{et al.} \cite{BTM}.
This is a relatively new approach and it is not part of the standard \textsc{TALYS1.95} distribution. \\

The OMP by Avrigeanu \textit{et al.} also uses a real part derived from microscopic nucleon-nucleon interactions to reduce the number of free parameters from nine to six when fitting to experimental data.
After the parameters used in the imaginary part are set, another fit is performed to obtain a set of three parameters for a phenomenological real part of Woods-Saxon shape \cite{aomp-avrigeanu_2010}.
This model is also used in the last set of theoretical values, that agrees very well with the experimental data stemming from the \textsc{Tendl-2019} database \cite{TENDL-2019}.
Within the considered energy range the results stemming from the \textsc{Talys1.95} calculation using the OMP by Avrigeanu \textit{et al.} with default values (light-green solid) for all other parameters yields virtually the same cross sections as those published in the \textsc{Tendl-2019} database (orange dash-dotted).
Any input parameters used in the \textsc{Tendl-2019} calculation that differ from \textsc{Talys1.95} default values therefore do not seem to have a drastic effect on the cross sections. \\

Cross section ratios predicted by \textsc{Talys1.95} calculations are shown in Fig. \ref{fig:ratio}.
For this the $\alpha$-OMP by Avrigeanu \textit{et al.} and the dispersive model by Demetriou \textit{et al.} were used.
Both models were chosen, because they have been found to reproduce experimental results at sub-Coulomb energies within a wide range of masses very well in previous publications \cite{Scholz_2020, Scholz_2016_108Cd, Ornelas_2015, Netterdon_2015_112Sn, Gyuerky2012}.
In addition, out of the five $\alpha$-OMPs capable of accurately reproducing the experimental results, these two models represent the highest and lowest predictions, respectively.
This is also true for the $^{164,166}$Yb($\alpha,\gamma$)$^{168,170}$Hf key reactions.
Both models are capable of accurately reproducing the cross section ratios. \\

By limiting the nuclear physics input to the five $\alpha$-OMPs found to reproduce the $^{168,170,172}$Yb($\alpha,n$) reaction cross sections well, uncertainties in reaction rate calculations can be reduced significantly. \textsc{Talys1.95} predictions for $^{164,166}$Yb($\alpha,\gamma$) reaction rates at T$_9$ = 3 based on these $\alpha$-OMPs and using default values for all other input parameters agree within a factor of five.
When considering all $\alpha$-OMPs available in \textsc{Talys1.95} reaction rate predictions vary by three orders of magnitude.
While this exemplifies the importance of an experimental database, it should be kept in mind that the extrapolation to unmeasured reactions relies on the predictive power of the models used.
Therefore, models should be globally applicable, simple, and based on microscopic physics where possible.
A deeper analysis would have to be based on a much broader set of data and is beyond the scope of this work.

\section{Summary and Conclusion}

A new target chamber for stacked-target experiments was designed providing an opportunity to investigate tiny cross sections within a reasonable experimental time frame.
It was used to obtain seven cross sections for the $^{170}$Yb($\alpha, n$)$^{173}$Hf reaction and six for the $^{172}$Yb($\alpha, n$)$^{175}$Hf reaction, which have not been measured before.
One of the major challenges in using the stacked-target method, namely calculating the interaction energy inside of each target and taking the uncertainties in the determination of target thicknesses into account, was solved by using a \textsc{GEANT4} simulation.
The successful application of a background subtraction improved the sensitivity limit, reached in this experiment, by a factor of four.
In addition, an upper limit for the reaction cross section at 12.3 MeV could be deduced for the $^{172}$Yb($\alpha, n$)$^{175}$Hf reaction.
All measured cross sections were validated using the $^{55}$Mn($\alpha, (2)n$)$^{57,58}$Co reference reactions.
Five different $\alpha$-OMPs capable of accurately predicting the cross sections without requiering any adjustments to the default parameters were identified.
The results obtained in this work were used to calculate cross section ratios for ($\alpha, n$) reactions on the three lowest mass, stable, even-even nuclei in the ytterbium chain.
These ratios can be reproduced by theory as well.
To complete the picture for all stable even-even nuclei in the ytterbium chain ($\alpha, n$) cross sections for $^{174,176}$Yb are needed.
Since the products of both reactions are stable, this requires in-beam experiments. \\

\begin{acknowledgments}
We gratefully thank K. O. Zell and A. Blazhev for the target preparation, H. W. Becker and V. Foteinou of the Ruhr-Universit\"at Bochum for their assistance during the RBS measurements, U. Giesen of the Physikalisch Technische Bundesanstalt Braunschweig for lending us parts of the target chamber and S. Thiel for adapting these parts to our needs.
This project has been supported by the Deutsche Forschungsgemeinschaft under the contract ZI 510/8-2 and by the European Union (ChETEC-INFRA, project no. 101008324)
\end{acknowledgments}

%

\end{document}